\documentclass[useAMS,usenatbib]{mn2e}

\usepackage[dvips]{graphicx}
\usepackage{color}
\usepackage{wrapfig}
\usepackage{amssymb}
\newcommand{\lseq}{\mbox{\raisebox{-0.7ex}{$\;\stackrel{<}{\sim}\;$}}}
\newcommand{\gseq}{\mbox{\raisebox{-0.7ex}{$\;\stackrel{>}{\sim}\;$}}}

\newcommand{\Msun}{M_{\odot}}
\newcommand{\Mh}{M_h}
\newcommand{\fesc}{f_{\rm esc}}
\newcommand{\rhoth}{\rho_{\rm th}}
\newcommand{\Nion}{\dot{N}_{\rm ion}}
\newcommand{\Rrel}{R_{\rm rel}}

\title[Escape of ionizing photons]
{Escape fraction of ionizing photons from high-redshift galaxies in cosmological SPH simulations}
\author[Yajima et al.]
{Hidenobu Yajima$^{1}$\thanks{E-mail: yuh19@psu.edu (HY);
},
Jun-Hwan Choi$^{2}$,
and Kentaro Nagamine$^{2}$
\\
$^{1}$Department of Astronomy and Astrophysics, Pennsylvania State University,
525 Davey Lab, University Park, PA 16802, USA\\
$^{2}$Department of Physics and Astronomy, University of Nevada, Las Vegas,
4505 S. Maryland Pkwy, Las Vegas, NV, 89154-4002, U.S.A.
}

\begin{document}


\pagerange{\pageref{firstpage}--\pageref{lastpage}} \pubyear{2008}

\maketitle

\label{firstpage}

%
%
\begin{abstract}

Combing the three-dimensional radiative transfer (RT) calculation and cosmological SPH simulations, we study the escape fraction of ionizing photons ($\fesc$) of high-redshift galaxies at $z=3-6$.
Our simulations cover the halo mass range of $\Mh = 10^{9} - 10^{12} \Msun$.
We postprocess several hundred simulated galaxies with the Authentic Radiative Transfer (ART) code to study the halo mass dependence of $\fesc$.
In this paper, we restrict ourselves to the transfer of stellar radiation
from local stellar population in each dark matter halo.
We find that the average $\fesc$ steeply decreases as the halo mass increases,
with a large scatter for the lower mass haloes.
The low mass haloes with $\Mh \sim 10^{9} \Msun$ have large values of
$\fesc$ (with an average of $\sim 0.4$), whereas
the massive haloes with $\Mh \sim 10^{11} \Msun$ show small values of
$\fesc$ (with an average of $\sim 0.07$).
This is because in our simulations, the massive haloes show more clumpy
structure in gas distribution, and star-forming regions are embedded inside these clumps, making it more difficult for the ionizing photons to escape.
On the other hand, in low mass haloes, there are often conical regions of highly ionized gas due to the shifted location of young star clusters from the center of dark matter halo, which allows the ionizing photons to escape more easily than in the high-mass haloes. 
By counting the number of escaped ionizing photons, we show that
the star-forming galaxies can ionize the intergalactic medium at $z=3-6$.
The main contributor to the ionizing photons is the haloes with $\Mh \lesssim 10^{10}\,\Msun$ owing to their high $\fesc$. 
The large dispersion in $\fesc$ suggests that there may be various sizes of H{\sc ii} bubbles around the haloes even with the same mass in the early stages of reionization. 
We also examine the effect of UV background radiation field on $\fesc$ using simple, four different treatment of UV background.
\end{abstract}

%
%

%
%
\section{INTRODUCTION}
\label{sec:intro}

Observations of cosmic microwave background radiation provides
a wealth of information on the cosmic reionization history \citep[e.g.,][]{Page07, Dunkley09}.  For example, \citet{Komatsu10} showed that the reionization occurred at $z\sim 10.5$ assuming an instantaneous reionization scenario.
However, the detailed history of reionization and the nature of ionizing
sources are not yet fully understood.
Since the UV background (UVB) radiation can heat up the interstellar medium
(ISM) to $\sim 10^{4}$\,K and disturb star formation, UVB coupled with
the ionization history of the universe significantly influences the
galaxy formation \citep[e.g.,][]{Susa00,Umemura01,Susa04, Okamoto08a, Hasegawa09b}.
Therefore it is very important to study the UVB intensity and the nature of
ionizing sources.

\citet{Haardt96} pointed out that the UVB is dominated by quasars at $z<4$.
Using the SDSS sample, \citet{Fan01a} showed that the bright-end slope
of the quasar luminosity function at $z\gseq4$ are considerably steeper
than that at lower redshifts, and 
concluded that the quasars cannot maintain the ionization of IGM at $z\gseq4$.
Subsequently, much argument have been focused on the possibility that
the IGM is ionized mainly by the UV radiation from high-redshift (hereafter high-$z$) star-forming galaxies \citep[e.g.,][]{Fan06b,Bouwens07,Gnedin08b}.
The key quantity in determining the IGM ionization rate is the
escape fraction of ionizing photons \citep[e.g.,][]{Razoumov06b, Gnedin08a},
which is the number ratio of photons escaping from a galaxy to the intrinsically radiated photons by stars.  This parameter controls the contribution
to the UVB intensity from star-forming galaxies.
In this work, we examine the values of $\fesc$ in high-$z$
star-forming galaxies.

There are several observational constraints on $\fesc$ at $z \sim 3$.
\citet{Steidel01} found $f_{\rm esc,rel} \gseq  0.5 $ from the composite
spectrum of 29 Lyman Break Galaxies (LBGs) at $z\sim 3$,
where $f_{\rm esc,rel}$ is the {\it relative} fraction of escaping
Lyman continuum (900\,\AA) photons relative to the fraction of
escaping non-ionizing UV (1500\,\AA) photons.
It is usually defined as
\begin{equation}
f_{\rm esc,rel} \equiv \frac{(L1500/L900)_{\rm int}}{(F1500/F900)_{\rm obs}}
\exp(\tau^{\rm IGM}_{900}), \label{f_esc_rel}
\label{eq:fesc_rel}
\end{equation}
where $(F1500/F900)_{\rm obs}$, $(L1500/L900)_{\rm int}$ and $\tau^{\rm IGM}_{900}$
represent the observed 1500\,\AA/900\,\AA~ flux density ratio,
the intrinsic 1500\,\AA/900\,\AA~ luminosity density ratio, and
the line-of-sight opacity of the IGM for 900\,\AA~ photons, respectively.
Equation~(\ref{eq:fesc_rel}) compares the observed flux density ratio
(corrected for the IGM opacity) with the models of UV spectral energy
distribution of star-forming galaxies.

\citet{Giallongo02} and \citet{Inoue05} estimated the upper limit of
$f_{\rm esc,rel} \lseq 0.1-0.4$ for some LBGs at $z\sim 3$.
\citet{Shapley06} directly detected the escaped 
ionizing photons from 2 LBGs in the SSA22 field at $z=3.1$, 
and estimated the average value of $f_{\rm esc,rel}= 0.14$.
Moreover, \citet{Iwata09} successfully detected the Lyman continuum emission
from 10 Ly-$\alpha$ emitters (LAEs) and 7 LBGs within a sample of 198 LAEs and LBGs in the SSA22 field.
They showed that the mean value of $f_{\rm esc,rel}$ for the 7 LBGs is 0.11 after correcting for dust extinction, and 0.20 if the IGM absorption is taken into account.

In the early theoretical works, some authors studied the $\fesc$ with 
ideally modelled galaxies.  
For example, \citet{Dove94} estimated the $\fesc$ of Milky Way type galaxy
using a semi-analytic method, and reported $\fesc \sim 0.07$.
\citet{Ricotti00} investigated the dependence of $\fesc$ on various 
physical quantities, such as the collapse redshift and  
star formation efficiency using a semi-analytic method.
\citet{Wood00} and \citet{Ciardi02} studied the effect of 
inhomogeneous structure of gas on $\fesc$, and showed that 
$\fesc$ increases in clumpy systems by a factor of $>2$ 
than in a homogeneous gas distribution.
\citet{Dove00} investigated the influence of bubbles made by 
supernovae on $\fesc$ using a semi-analytic method.
Using numerical simulations, \citet{Fujita03} studied 
the effect of supernovae feedback, and reported a high $\fesc$ ($> 0.2$) 
for a disk galaxy with $\Mh = 10^{8} - 10^{10} \Msun$. 

Theoretical studies in a more fully cosmological environment can be performed 
by combining cosmological hydrodynamic simulations of galaxy formation and
a three-dimensional radiative transfer calculation. 
For example, \citet[][hereafter Y09]{Yajima09} post-processed the
eulerian hydrodynamic simulation of \citet{Mori06}
with RT, and showed that the galaxies in an isolated halo of $\Mh =10^{11} \Msun$ can have relatively large values of $\fesc = 0.17 - 0.47$.
Moreover they found that $\fesc$ decreases gradually as a function of time
owing to the dust pollution and the shifting star formation sites. 

On the other hand, \citet{Razoumov06b} examined the escape fractions
of two galaxies in a cosmological SPH simulation from $z=3.8$ to 2.4, which later become Milky Way type disk galaxies at $z=0$.
They found small values of $\fesc < 0.1$, in disagreement 
with Y09.  However they also reported that $\fesc$ decreases with redshift from $z=3.8$ to 2.4, in qualitative agreement with Y09.

\citet{Razoumov10} further examined the $\fesc$ of star-forming galaxies
in a wide mass range ($\Mh = 10^{7.8} - 10^{11.5} \Msun$) at $z = 4 - 10$,
and found that $\fesc$ decreases steeply as the halo mass increases in their cosmological SPH simulations, in contrast to the work by \citet{Gnedin08a} and \citet{Wise09}.

Using cosmological AMR simulations, \citet{Gnedin08a} reported that haloes with $\Mh = 10^{11} - 10^{12} \Msun$ have $\fesc = 0.01 - 0.03$, and much lower $\fesc$ for lower mass haloes with $\Mh = 10^{10} - 10^{11} \Msun$. Their results suggest that $\fesc$ increases with halo mass, at least in the range of $\Mh = 10^{10} - 10^{11} \Msun$.

\citet{Wise09} extracted 10 haloes with masses $\Mh = 3\times 10^6 - 3\times 10^9 \Msun$ at $z=8$ from cosmological AMR radiation hydrodynamic simulations, and examined the escape fraction of ionizing photons.
They found that $\fesc$ fluctuates rapidly on a time-scale of a few to 10 Myrs depending on the star formation rates, and varies widely from almost zero to nearly unity.  They found $\fesc \sim 0.4$ for a normal IMF for the haloes with $\Mh = 10^{7.5} - 10^{9.5} \Msun$, but $\fesc = 0.05 - 0.1$ for lower mass haloes, disregarding the effect of dust.

Although the halo mass dependence of $\fesc$ is very important for the study of cosmic reionization, there are significant differences in the theoretical estimates from cosmological hydrodynamic simulations as described above. In particular, in these previous works, the number of studied haloes has been very small ($\sim 10$), therefore it has been difficult to gauge the halo mass dependence of $\fesc$.    In the present paper, we calculate the values of $\fesc$ for a much larger number of haloes (several hundreds) in cosmological volumes of comoving 10$-$100\,$h^{-1}$Mpc, and examine its halo mass dependence.  In addition, we study the effects of interstellar dust and UVB radiation on $\fesc$.

The outline of the paper is as follows.
In \S~\ref{sec:model}, the models and numerical methods are described.
We present the results on escape fractions in \S~\ref{sec:result},
and discuss the dust effect and the contribution of star-forming galaxies
to the reionization of the universe in \S~\ref{sec:discussion}.
We then summarise in \S~\ref{sec:summary}.

%
%
\section[]{MODEL AND METHOD}
\label{sec:model}

\begin{table*}
\begin{center}
\begin{tabular}{ccccccc}
\hline
Series &  Box-size & ${N_{\rm p}}$ & $m_{\rm DM}$ & $m_{\rm gas}$ & $\epsilon$ & $z_{\rm end}$ \\
\hline
N144L10      & 10.00   & $2 \times 144^3$ & $1.97 \times 10^7$ & $4.04 \times 10^6$ & 2.78 & 2.75 \\
\hline
N216L10    & 10.0  & $2\times 216^3$  & $5.96 \times 10^6$ & $1.21 \times 10^6$ &  1.85  & 2.75 \cr
\hline
N400L100   & 100.0 & $2\times 400^3$  & $9.12 \times 10^8$ & $1.91 \times 10^8$ &  6.45  & 0.0 \cr
\hline
\end{tabular}
\caption{
Series of simulations employed for the present study.
The box-size is given in units of $h^{-1}$Mpc, ${N_{\rm p}}$ is the particle
number of dark matter and gas (hence $\times\, 2$), $m_{\rm DM}$ and
$m_{\rm gas}$ are the masses of dark matter and gas particles in units of
$h^{-1}M_{\sun}$, respectively, $\epsilon$ is the comoving gravitational
softening length in units of $h^{-1}$kpc, and $z_{\rm end}$ is the ending
redshift of the simulation.  The value of $\epsilon$ is a measure of
spatial resolution.
}
\label{table:sim}
\end{center}
\end{table*}

\subsection{Simulations}

We use an updated and modified version of the Tree-particle-mesh
(TreePM) smoothed particle hydrodynamics (SPH) code GADGET-3
\citep[originally described in][]{Springel05e}.
The SPH calculation is performed based on the entropy
conservative formulation \citep{Springel02}.
Our fiducial code includes radiative cooling by H, He, and
metals \citep{Choi09a}, star formation, supernova feedback,
a phenomenological model for galactic winds \citep{Choi10}, 
and a sub-resolution model of multiphase ISM \citep{Springel03a}.
We also include the heating by a uniform UVB, which we will
discuss more in \S~\ref{sec:UVB}.

In this multiphase ISM model, high-density ISM is pictured to be a
two-phase fluid consisting of cold clouds in pressure equilibrium with
a hot ambient phase.  Cold clouds grow by radiative cooling out of the
hot medium, and this material forms the reservoir of baryons available
for star formation. The star formation rate (SFR) is estimated for each gas
particle that have densities above the threshold density, and
the star particles are spawned statistically based on the SFR.
For the star formation model, the ``Pressure model'' described in 
\citet{Choi09b} is being used.
This model estimates the SFR based on the local gas pressure rather than
the gas density, and implicitly considers the effect of H$_2$ formation.

The simulations used in this paper also uses the new ``Multicomponent
Variable Velocity'' wind model developed by \citet{Choi10}.
This new wind model is based on both the energy-driven wind and 
the momentum-driven wind model discussed by \citet{Murray05},  
and the wind speed in this model depends on the galaxy stellar mass and SFR.
It gives more favorable results, and agrees better with the observations of, e.g., cosmic C\,{\sc iv} mass density and IGM temperature than the previous model with a constant wind speed.
To enable this new wind model, \citet{Choi10} implemented a on-the-fly group-finder into GADGET-3 to compute the galaxy masses and SFRs while the simulation is running.  The group-finder, which is a simplified variant of the \textup{SUBFIND} algorithm developed by \citet{Springel01}, 
identifies the isolated groups of star and gas particles (i.e., galaxies) based on the baryonic density field. The detailed procedure of this galaxy grouping is described in \citet{Nag04e}.
The outer baryonic density threshold for a galaxy is 0.01$\rhoth$, where $\rhoth$ is the threshold density for star formation.

The parameters of the simulations are summarised in Table~\ref{table:sim}.
Due to the computational load of the RT calculation, we primarily use
the N144L10 series for our main results.
The other two series (N216L10 and N400L100) are used for the resolution tests, with a fewer number of simulated galaxies postprocessed with RT. 
The adopted cosmological parameters are consistent with the WMAP results:
$H_0 = 72$\,km\,s$^{-1}$\,Mpc$^{-1}$ ($h=0.72$),
$\Omega_M=0.26$, $\Omega_{\Lambda}=0.74$, $\Omega_b=0.044$,
$\sigma_8 = 0.80$, and $n_s = 0.96$.


\subsection{Stellar Radiation Transfer}
\label{sec:radiation}

To estimate $\fesc$, we compute the stellar radiation transfer
and the ionization structure of gas in each dark matter halo
by post-processing the simulation output.
First we set up a uniform grid around each dark matter halo
with a grid cell size equal
to the gravitational softening length of the simulation, and translate
the SPH gas information into a gridded data.
The grid typically have $\sim$300$^3$ cells for
high-mass haloes ($\sim 10^{12} \Msun$) and $\sim$70$^3$ cells
for low-mass haloes ($\sim 10^{9} \Msun$).

The RT scheme used in this paper is the Authentic Radiation Transfer (ART) method originally developed by \citet{Nakamoto01}, and the treatment is basically the same as in Y09.  The performance of our RT code has already been reported as part of the RT code comparison study by \citet{Iliev06}, and it can calculate the ionization structure precisely \citep[see Figure 6 in][]{Iliev06}.

Usually the {\it short-characteristic} method is computationally cheaper
than the {\it long-characteristic} method by a factor of $N_{\rm g}$, 
where $N_{\rm g}$ is the grid size of RT calculation.
In the short-characteristic method, the amount of calculation is reduced by the interpolation of optical depth from the nearest grids.
However the short-characteristic method suffers from an artificial photon diffusion effect.  Our ART method is based on the long-characteristic method, and it is devised to reduce the calculation amount to a similar level as the short-characteristic method.  Hence the ART method is suitable for the calculation of $\fesc$ for a large number of galaxies.  As for the scattering of photons, we employ the on-the-spot approximation \citep{Osterbrock89}, in which the scattered photons are assumed to be absorbed immediately on the spot.

In this work, the RT equation is solved along 
$N_{\rm g}^{2}$ rays with uniform angular resolution from each source.
The number of ionizing photons emitted from the source stars
is computed based on the theoretical spectral energy distribution (SED)
given by the population synthesis code P\'{E}GASE v2.0 \citep{Fioc97}.
We take only the star particles that are younger than $10^{7}$ yrs as the sources of ionizing photons, and consider the effect of age and metallicty of the stellar population by interpolating the table generated from the result of P\'{E}GASE.  We shoot the radiation rays in a radial fashion from each star particles. 
We assume the \citet{Salpeter55} initial mass function with the mass range of $0.1 - 50 \Msun$.

Typically the postprocessing RT calculation takes about 100 hours for a large grid of 300$^3$, and 1 hour for a small grid of 70$^3$ on a single CPU.
Our ART code is parallelized by MPI, and has a high parallelization efficiency.
We process each star particle in parallel, and each CPU calculates the radiation field from each star particle. In practice, we use $\sim 1-128$ CPUs simultaneously to process one halo with RT.


\subsection{Dust Attenuation}
\label{sec:dust}

We also include the effect of dust attenuation by distributing
the interstellar dust proportionally to the metallicity,
with a size distribution of $n_{\rm d}(a_{\rm d}) \propto a^{-3.5}_{\rm d}$
\citep{Mathis77}, where $a_{\rm d}$ is the radius of a dust grain.
We adopt the dust grain size range of $0.1 -1.0\,\mu$m as our fiducial model.
The dust mass is calculated as $m_{\rm d}=0.01m_{\rm g}(Z/Z_{\odot})$ \citep{Draine07}, where $m_{\rm d}$, $m_{\rm g}$, and $Z$ are the dust mass, gas mass, and metallicity in a grid.
The density of a dust grain is assumed to be 3\,g\,cm$^{-3}$ like silicates.
The dust opacity is given by
$d\tau _{\rm dust} = Q(\nu) \pi a^{2}_{\rm d} n_{\rm d} d\ell$,
where $Q(\nu), a_{\rm d}, n_{\rm d}$, and $d\ell$ are the absorption efficiency factor, dust size, number density of dust grains, and path length, respectively.
Since the assumed range of dust size is larger than the wavelength of Lyman limit, we assume $Q(\nu)=1$ for ionizing photons \citep{DL84}.

\begin{figure*}
\begin{center}
\includegraphics[angle=270,scale=0.6]{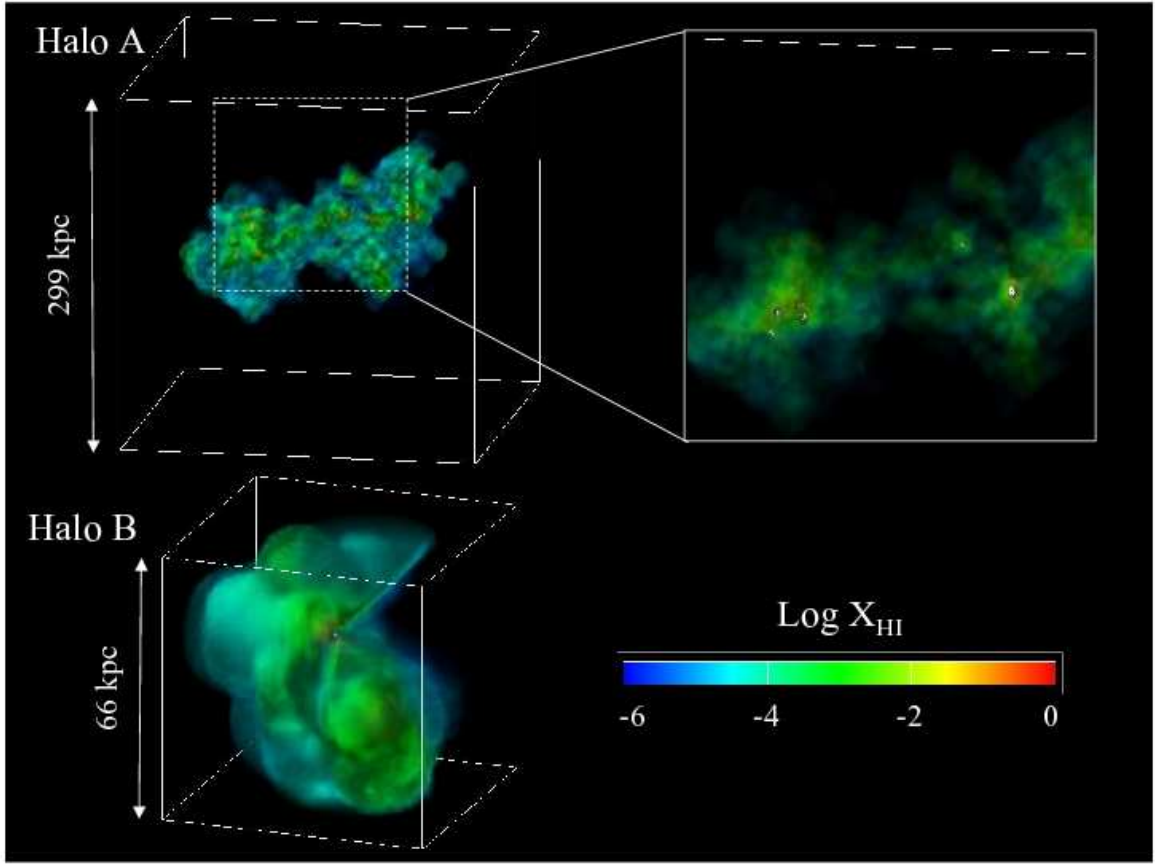}
\caption{
$Upper$:The ionization structure of Halo A($\Mh \sim 7 \times 10^{11} \Msun$).
$lower$:The ionization structure of Halo B($\Mh \sim 1 \times 10^{10} \Msun$).
Color indicates the neutral fraction of hydrogen in log scale.
White points show the positions of young star clusters.
}
\label{fig:ionize}
\end{center}
\end{figure*}


\subsection{UV Background Radiation}
\label{sec:UVB}

The baryonic gas in galaxies can be ionized by the UVB, and heated up to $\sim 10^4$\,K.
It would be ideal to compute the RT of UVB as well as the stellar radiation,
but in practice it is a very expensive calculation.

Let us briefly explain why the UVB RT calculation is much heavier than the stellar radiation transfer. 
For stellar radiation, using on-the-spot approximation, we only calculate the RT along the angular rays between stars and grids with the dilution factor by the distance, whereas for UVB we have to calculate the RT of all angular rays. 
Therefore in the ART method, the number of rays that we have to calculate is
$N_{\rm star} \times N_\phi \times N_\theta \times N_{\rm path} \approx N_{\rm star} \times N_{\rm g}^{3}$ for stellar radiation, and $N_{\rm g}^{5}$ for UVB.
If $N_{\rm g}^{2} > N_{\rm star} $, the calculation amount for UVB is larger than the stellar radiation.
In practice, $N_{\rm g}^{2}$ is greater than $N_{\rm star}$ by $\sim 2-4$ orders of magnitude.

Due to this difficulty of UVB RT, usually a uniform UVB radiation field
with an optically thin approximation is assumed across the simulation box
as a simple approximation. 
Our fiducial simulation also includes a uniform UVB with a modified
\citet{Haardt96} spectrum \citep[see][]{Dave99}, where the reionization
takes place at $z\simeq 6$ as suggested by the quasar observations
\citep[e.g.,][]{Becker01} and stellar radiative transfer calculations
\citep[e.g.,][]{Sokasian03}.

However, the optically thin approximation is very crude, and the effects of
different UVB has not been explored very much.
Here we use following four N144L10 simulations with different treatment of
UVB to examine the effects of UVB:
\begin{enumerate}
\item {\bf Fiducial}:
A uniform UVB radiation field with an optically thin approximation is assumed as stated above. \\
\item {\bf MH0.5} (modified Haardt 0.5): 
The ISM is optically thin to the same UVB, however the intensity of UVB is reduced to the half of the Fiducial run. \\
\item {\bf OTUV} (optically thick UV):
The ISM is optically thin to the UVB in the lower density regions
with $n_{\rm H} < 0.01 \rhoth = 6.34 \times 10^{-3}$\,cm$^{-3}$, but completely
optically thick in higher density region ($n_{\rm H} \ge 0.01 \rhoth$),
where $\rhoth$ is the threshold density, above which star formation is allowed.
The value of $\rhoth$ was determined by \cite{Choi09b} based
on the observed SF cut-off column density of the Kennicutt law.
The OTUV method implicitly assumes that the UVB cannot penetrate into
the high density regions by self-shielding.
We find that this treatment reproduces the observed H\,{\sc i} column density distribution function very well \citep{Nagamine10}, and more detailed analyses using RT calculation supports this self-shielding density (Yajima et al. 2010, in preparation). \\
\item {\bf no-UVB}: UVB does not exist at all.
\end{enumerate}

\vspace{0.2cm}
We compute the ionization structure in each halo 
by solving the equation of ionization equilibrium as follows:
\begin{equation}
(\Gamma_{\rm UVB}^{\gamma}+ \Gamma_{\rm star}^{\gamma}) \; n_{\rm HI}
+ \Gamma^{\rm C}\; n_{\rm HI} \; n_{\rm e} = \alpha_{\rm B}\; n_{\rm HII} \; n_{\rm e},
\end{equation}
where $\Gamma_{\rm UVB}^{\gamma}$ and $\Gamma_{\rm star}^{\gamma}$
are the photoionization rate by UVB and stellar radiation;
$\Gamma^{\rm C}$ is the collisional ionization rate;
$n_{\rm e}, n_{\rm HI}$ and $n_{\rm HII}$ are the number density of
free-electron, neutral and ionized hydrogen, respectively;
$\alpha_{\rm B}$ is the total recombination coefficient to all bound excitation levels.
The value of $\Gamma_{\rm star}^{\gamma}$ is estimated by the full RT calculation,
but $\Gamma_{\rm UVB}^{\gamma}$ is computed with the optically thin approximation.

%
%
\section[]{RESULTS}
\label{sec:result}

\subsection{Ionization Structure}

Figure~\ref{fig:ionize} shows the ionization structure of gas in a high-mass 
halo (Halo ``A'') and a low-mass halo (Halo ``B'') in the N144L10 Fiducial
UVB run. 

The gas in Halo-A shows very complex and clumpy structure, going through 
continuous merging processes. 
The young star clusters are born in dense, neutral 
clumpy regions and irradiate the ambient ISM. 
However, the dense neutral gas clumps survive owing to high 
recombination rates.

On the other hand, the Halo-B shows more or less spherical gas distribution 
before we process it with RT.  Star clusters are born near the central 
high-density region.
Once the halo is processed with RT, most of the low density gas on the
right-hand-side is ionized by the UVB and stellar radiation.
In particular, when the location of a young star cluster is slightly
off-center, it can ionize one side of the halo preferentially, and creates
a conical region of highly ionized region.
The high density gas on the left-hand-side of the star cluster remains 
neutral, and the ionizing photons cannot escape to the left-hand-side. 
The value of angle-averaged $\fesc$ of each halo is basically determined 
by the covering fraction of these highly ionized region.


\subsection{Escape Fraction}

We estimate the average value of $\fesc$ for each dark matter halo as follows. 
For each light ray from each star particle in the halo, we count up the number of escaped ionizing photons by integrating the transferred spectrum as a function of wavelength, and then divide it by the total number of intrinsically radiated ionizing photons.  
Then the values of $\fesc$ are averaged over all the rays coming out from the halo at the surface of the grid that was set up around the halo. 
Hereafter $\fesc$ denotes the angle-averaged value for each halo most of the time. 

\subsubsection{Halo Mass and Redshift Dependence}
\label{sec:dependence}

\begin{figure}
\begin{center}
\includegraphics[scale=0.4]{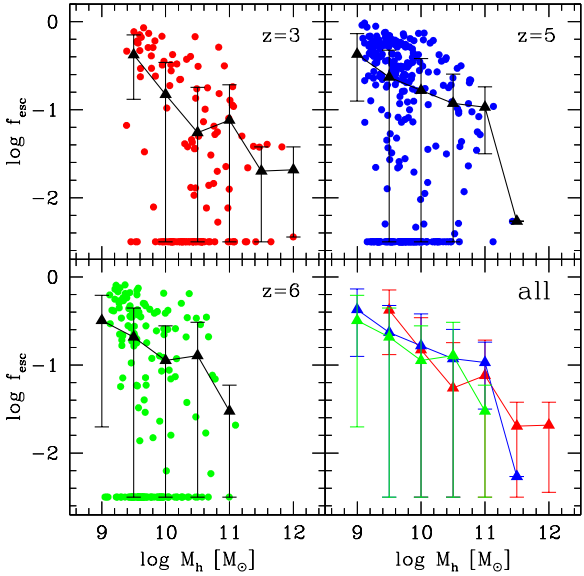}
\caption{
Escape fraction as a function of halo mass at $z=3-6$ for 
the N144L10 Fiducial UVB run.  Different colors are used for 
different redshifts (red: $z=3$, blue: $z=5$, green: $z=6$).
The triangles in the bottom right panel show the mean values in each 
mass bin with 1-$\sigma$ error bars.
The data points with $\log \fesc < -2.5$ are shown at $\log \fesc = -2.5$ 
for plotting purposes.
}
\label{fig:redshift}
\end{center}
\end{figure}

\begin{figure}
\begin{center}
\includegraphics[scale=0.42]{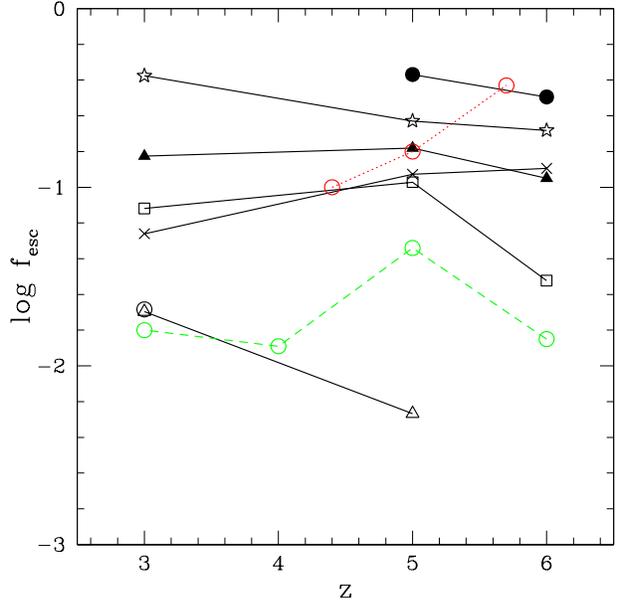}
\caption{
Mean escape fraction in each mass bin of Fig.~\ref{fig:redshift} 
as a function of redshift. Different symbols indicate different 
halo mass ranges (filled circles: $10^{8.75} - 10^{9.25} \Msun$, 
stars: $10^{9.25} - 10^{9.75} \Msun$, 
filled triangles: $10^{9.75} - 10^{10.25} \Msun$, 
open squares: $10^{10.25} - 10^{10.75} \Msun$,
crosses: $10^{10.75} - 10^{11.25} \Msun$, 
open triangles: $10^{11.25} - 10^{11.75} \Msun$, 
and open circle: $10^{11.75} - 10^{12.25} \Msun$).
The green open circles are for a relatively massive halo 
($M_{\rm total} = 4 \times 10^{11} \Msun$ at $z=3$) examined by 
\citet{Gnedin08a}, which is to be compared with our open triangles. 
The red open circles are for a similarly massive halo 
($\Mh \sim 10^{11} \Msun$) in the S33 run of \citet{Razoumov10}, 
which is to be compared with our crosses. 
}
\label{fig:mass_redshift}
\end{center}
\end{figure}

\begin{figure}
\begin{center}
\includegraphics[scale=0.4]{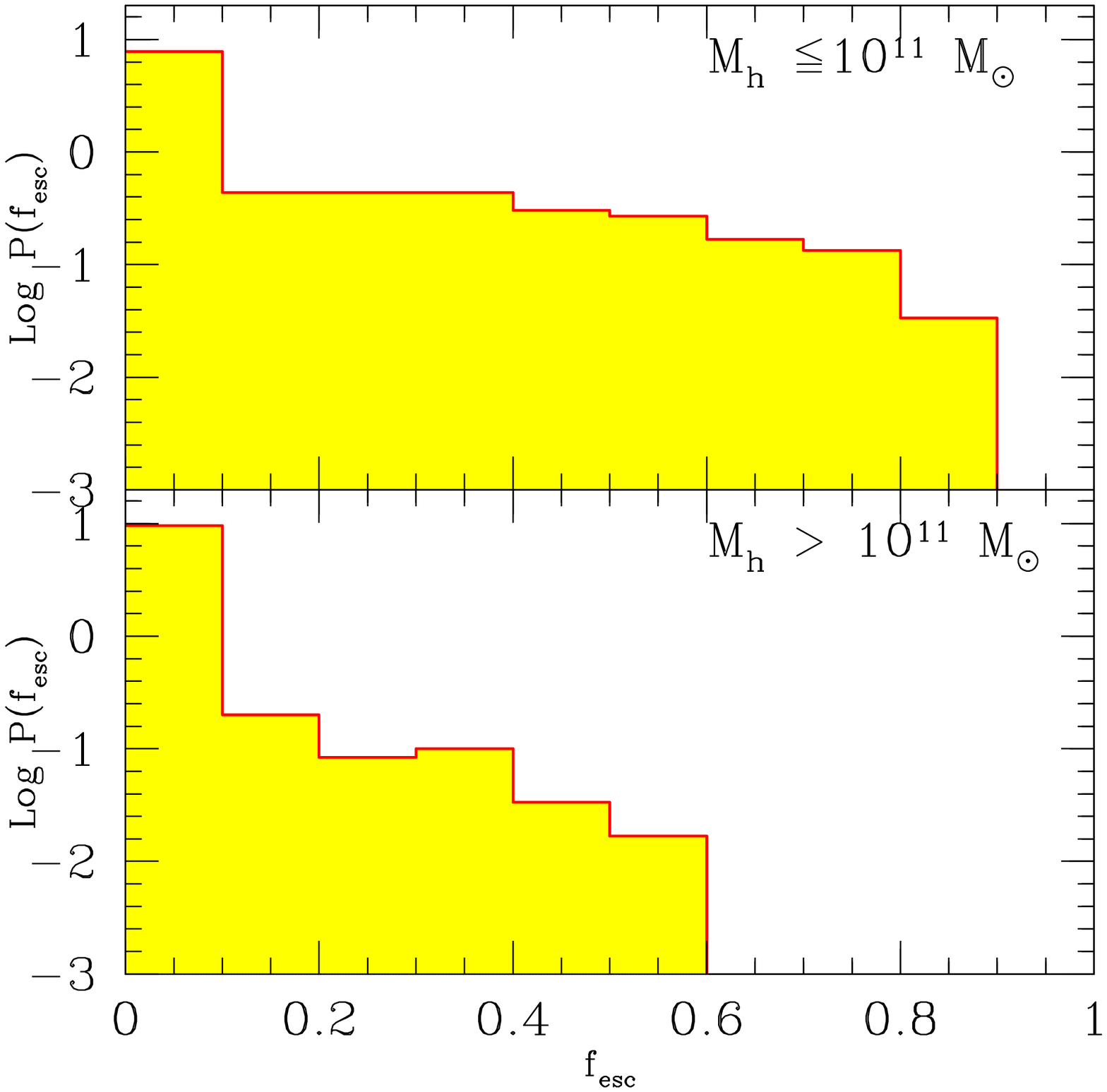}
\caption{
Probability distribution function (PDF) of 
star particles as a function of $\fesc$ at $z=3$ 
in the N144L10 Fiducial UVB run. 
The top panel shows the PDF of all star particles in haloes with 
$\Mh \le 10^{11} \Msun$,
and the bottom panel is for haloes with $\Mh > 10^{11} \Msun$.
Lower mass haloes contain more star particles with larger $\fesc$. 
}
\label{fig:pdf-mass-divided}
\end{center}
\end{figure}

Figure~\ref{fig:redshift} shows the $\fesc$ as a function of halo mass
at $z=3-6$ for the N144L10 Fiducial UVB run.
The solid triangles are the average values of $\fesc$ in each mass bin
with 1-$\sigma$ error bars. We first take the average in the linear scale, 
and then take the logarithm for the data points 
(i.e., $\log \langle \fesc \rangle$).   
At all redshifts, we find a clear qualitative trend that the mean $\fesc$
declines with increasing halo mass.

Our results are similar to the trend reported by \citet{Razoumov10},
although our $\fesc$ values are lower than theirs. 
Our results show the opposite trend to \citet{Gnedin08a},
although for high-mass galaxies, our $\fesc$ is similar to theirs. 
In \citet{Gnedin08a}, most of the simulated galaxies show a 
disk-like structure.  In their scenario, for low mass galaxies, 
stars are born after the disk is formed, and most young stars 
are embedded deep inside the disk.  As a result, most of the 
ionizing photons are absorbed in the disk, and $\fesc$ is low.  
However in more massive galaxies, some star-clusters can form  
near the edge of dense disk, and they can be exposed by the mergers
of galaxies. As a result of this effect, they argued that 
more ionizing photons can escape from higher mass haloes.
Therefore the geometry of simulated galaxies may cause the difference 
in the halo mass dependence of $\fesc$.  

The galaxy sample size in \citet{Gnedin08a} and \citet{Razoumov10}
is $\sim 10 - 20$, and it is somewhat small to discuss the systematic trend 
of $\fesc$ as a function of halo mass. 
The spatial resolution of \citet{Gnedin08a}, which is adaptively refined 
depending on the gas density, is from $\sim 17$\,kpc to $260$\,pc.
Although this resolution of maximum refinement level is better than that 
of ours, their RT scheme (OTVET) is coarser in estimating the 
ionization structure \citep{Iliev06}.
These differences in the accuracy and resolution of fluid and RT calculations may have caused the difference in $\fesc$.
We will further discuss the possible resolution effects in Section~\ref{sec:summary}. 

Figure~\ref{fig:mass_redshift} shows the redshift evolution of 
mean $\fesc$ in each halo mass bin.
In \citet{Razoumov10}, the $\fesc$ of high-mass haloes with 
$\Mh > 10^{10} \Msun$ clearly decreases with redshift (blue open circles), 
and that of the low-mass haloes does not change largely.
On the other hand, our results and \citet{Gnedin08a}
indicate that $\fesc$ of high-mass haloes with $\Mh > 10^{10} \Msun$ 
does not change largely with redshift.
For low-mass haloes with $\Mh < 10^{10} \Msun$, it seems that 
$\fesc$ is increasing slightly with decreasing redshift in our simulations. 
This might be due to the increasing cosmic SFR density and 
increasing UVB intensity from $z=6$ to $z=3$. 
Indeed, if we calculate the radiative transfer without the 
contribution of UVB in Eq.\,(2) for the Fiducial run at $z=3$ with the 
same gas and stellar distribution,  $\fesc$ decreases 
by $\sim 10-20$ per cent. 
In addition, the mass fraction of gas with $\log n_{\rm H}>0.6$ within haloes
increases with increasing redshift, which leads to lower escape fraction 
due to higher recombination rate. 

Figure~\ref{fig:pdf-mass-divided} shows the probability distribution 
function (PDF) of star particles as a function of 
$\fesc$ in haloes with $\Mh \le 10^{11} \Msun$ (top panel) and 
$\Mh > 10^{11} \Msun$ (bottom panel). 
The probability is defined by
$P(\fesc) = N_{\rm star}(\fesc \sim \fesc + \Delta \fesc) / ( N_{\rm star, total} \Delta \fesc)$,
where $N_{\rm star}$ is the number of star particles that have the value 
of $\fesc$, $N_{\rm star,total}$ is the total number of source star particles, 
and $\Delta \fesc$ is the bin width. 
The figure shows that the lower mass haloes have a longer tail 
towards higher values of $\fesc$.
Since the ionization structure in low-mass haloes shows conical regions 
of highly ionized gas, ionizing photons can escape easily through these 
ionized cones, but not through other angular directions covered by 
highly neutral gas.  This allows for some star particles in 
lower mass haloes to have high $\fesc$. 
On the other hand, the higher mass haloes show very complex and 
clumpy distribution of highly neutral gas, therefore it is 
more difficult for the ionizing photons to escape,
and there are no star particles with $\fesc > 0.6$.
Thus the PDF for higher mass haloes is concentrated at $\fesc < 0.1$.

We also find that there is a large dispersion in 
$\fesc$ for the low-mass haloes with $\Mh \le 10^{11} \Msun$.
This result may explain some of the recent observations.
For example, \citet{Shapley06} and \citet{Iwata09} detected 
ionizing radiation from high-$z$ galaxies, with a detection rate 
of about 10 per cent.
The detected galaxies show extremely high $\fesc$ ($\sim 100$ per cent).
Our results do not show such high values of $\fesc$,
however the $\fesc$ derived by \citet{Shapley06} and \citet{Iwata09}
are estimated from the flux ratio at the Lyman limit and UV continuum.
Recently \citet{Inoue10} pointed out that the nebulae emission lines 
can boost the above flux ratio, leading to a very high $\fesc$
with an assumption that the $\fesc$ of nebular and 
stellar emission is a few tens of per cent. 

In our simulation sample, about 10 per cent show high $\fesc$ ($> 0.4$).
These galaxies may corresponding to the recently observed objects with 
very high $\fesc$. 
Furthermore, \citet{Iwata09} showed that $\fesc$ decreases with 
increasing UV flux. 
In our simulations, SFR is positively correlated with halo mass, 
therefore our result of decreasing $\fesc$ with increasing halo mass 
is consistent with that of \citet{Iwata09}.


\subsubsection{Dependence on the UVB Models}

\begin{figure}
\begin{center}
\includegraphics[scale=0.42]{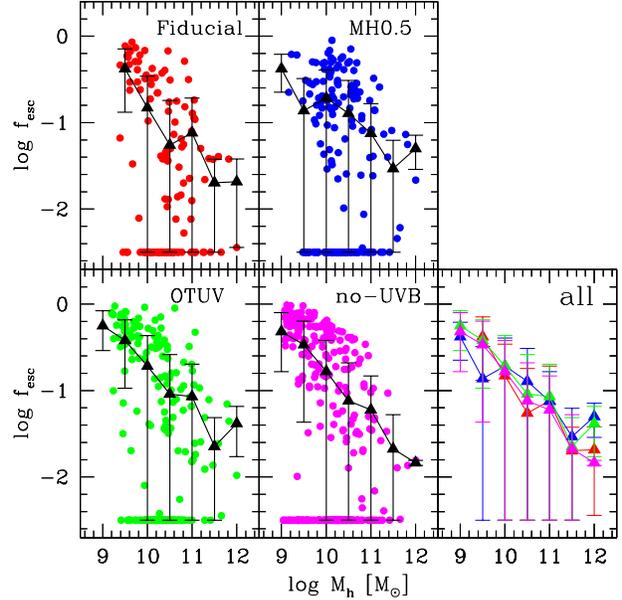}
\caption{
Escape fractions of ionizing photons ($\fesc$) for the four UVB models at $z=3$.
Different colors indicate different UVB models 
(red: Fiducial, blue: MH0.5, green: OTUV, and magenta: no-UVB). 
The triangles show the mean values in each mass bin with 1-$\sigma$ error bars.
The data points with $\log \fesc < -2.5$ are set to $\log \fesc = -2.5$ for plotting purposes.  There is not much differences between different UVB model runs due to self-regulation effect described in the text. 
}
\label{fig:uvall}
\end{center}
\end{figure}

Figure~\ref{fig:uvall} shows $\fesc$ as a function of halo masses for 
different UVB model runs at $z=3$.
Similarly to the Fiducial UVB model run, $\fesc$ decreases 
as the halo mass increases in all UVB models.
Since the ionization fraction of gas should increase with the increasing 
UVB intensity, one would naively expect a higher $\fesc$ for the runs with 
stronger UVB intensities. 
However, when the UVB intensity increases, the stronger gas heating results 
in less efficient cooling of gas and hence less star formation. 
This reduces the number of ionizing photons, and decreases the 
ionization fraction around the star-forming regions.  
Therefore these two competing effects 
counteract each other and self-regulate the ionization fraction, 
resulting in no significant differences in $\fesc$ between different 
UVB runs, as shown in the bottom right panel of Figure~\ref{fig:uvall}.

\begin{figure}
\includegraphics[scale=0.4]{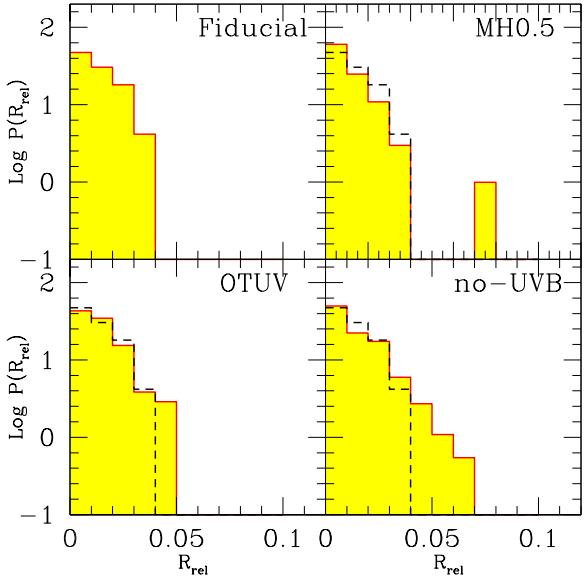}
\caption{ 
PDF of star particles as a function of $\Rrel$ in different UVB models at $z=3$, where $\Rrel$ is the relative distance between each star particle and the highest gas density peak in the halo, normalized by the maximum radius of the halo.
To focus on the scatter of $\fesc$, here we used only the low-mass haloes that include only one young star particle.
The result of the Fiducial UVB run is shown by the dashed histogram 
in other panels for comparison.  The runs with a weaker UVB have more young star particles at larger distances from the density peak. 
}
\label{fig:distance}
\end{figure}

We also considered the possibility that the sites of star formation 
might be different depending on the UVB intensity. 
Figure~\ref{fig:distance} shows the probability distribution function (PDF) 
of star particles as a function of relative distance ``$\Rrel$'' between each star particle and the highest gas density peak in the halo, normalized by the maximum radius of the halo ($\Rrel \equiv R_{\rm star} / R_{\rm halo}$).
To focus on the scatter of $\fesc$ that we see for the lower mass haloes, here we selected only the low-mass haloes that include only one young star particle.
(There could be other star particles that are older in the same halo.) 
This figure shows that, when the UVB intensity becomes weaker, 
star-forming regions are more spread out to larger radii.
Therefore the weaker UVB runs have more extended tails to larger radii 
compared to the Fiducial UVB model, progressively in the order of MH0.5, OTUV, and no-UVB run, although the probability in the extended tail is very small.
If a star cluster is farther away from the gas density peak,
the value of $\fesc$ would increase because the solid angle subtended 
by the high-density neutral clouds would become smaller.

\subsubsection{Origin of Scatter in $\fesc$}

\begin{figure}
\includegraphics[scale=0.45]{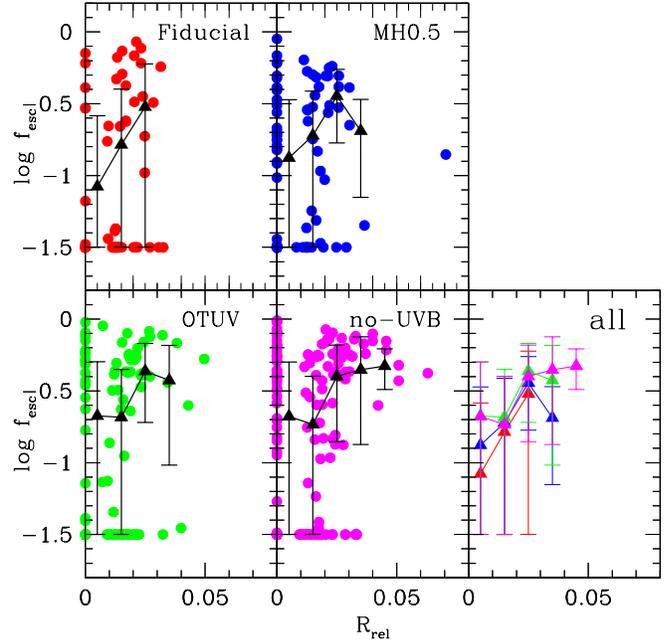}
\caption{
Escape fraction of the low-mass haloes as a function of $\Rrel$. 
Similarly to Figure~\ref{fig:distance}, only the low-mass haloes that include only one young star cluster are used. Different colors indicate different UVB models (red: Fiducial, blue: MH0.5, green: OTUV, and magenta: no-UVB). 
The triangles show the mean values in each $\Rrel$ bin with 1-$\sigma$ error bars.
The data points with $\log \fesc < -1.5$ are set to $\log \fesc = -1.5$ for plotting purposes.
}
\label{fig:fesc_distance}
\end{figure}

To further investigate the effect of star cluster locations on $\fesc$, 
we plot $\fesc$ as a function of $\Rrel$ for each halo in different UVB models 
in Figure~\ref{fig:fesc_distance}.
Here again, we selected only the low-mass haloes that include only one young star particle to focus on the scatter in $\fesc$ of the low-mass haloes.
The gas near the density peak has high recombination rate, and therefore optically thick.  The value of $\fesc$ strongly depends on the size of viewing angle to optically thick cloud from the source. 
As the source deviates from the central density peak, the viewing angle towards optically thick clouds decreases, and on average, $\fesc$ increases with increasing $\Rrel$. 
Moreover the mean value of $\fesc$ does not depend on the UVB models very much, although the scatter is somewhat larger at larger $\Rrel$ values owing to the small sampling.  At the lower values of $\Rrel$, the scatter among different UVB models are smaller, because UVB cannot penetrate into the high-density cloud by the self-shielding effect. 
The results shown in Figures~\ref{fig:distance} and \ref{fig:fesc_distance} suggest that the variation in $\Rrel$ is one of the key factors that determines the scatter in $\fesc$ for low-mass haloes.

\begin{figure}
\includegraphics[scale=0.45]{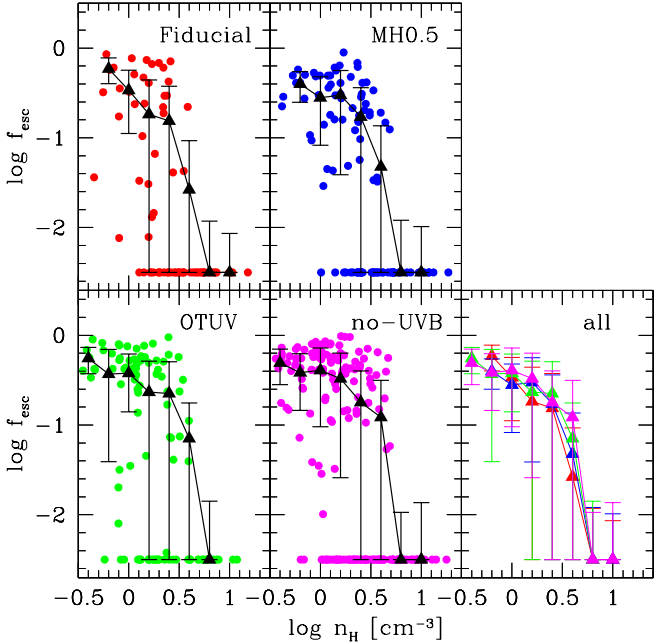}
\caption{
Escape fraction of the low-mass haloes as a function of hydrogen number density at the location of the star particle in different UVB models at $z=3$.
Similarly to Figure~\ref{fig:distance}, only the low-mass haloes that include only one young star cluster are used.
Different colors indicate different UVB models (red: Fiducial, blue: MH0.5, green: OTUV, and magenta: no-UVB). 
The triangles show the mean values in each mass bin with 1-$\sigma$ error bars.
The data points with $\log \fesc < -2.5$ are set to $\log \fesc = -2.5$ for plotting purposes.
}
\label{fig:fenh}
\end{figure}

Furthermore $\fesc$ may depend on the hydrogen number density at the location of star clusters, because the neutral hydrogen gas near the star clusters can effectively absorb ionizing photons.
Figure~\ref{fig:fenh} shows the $\fesc$ of low-mass haloes as a function of hydrogen number density $n_{\rm H}$ at the location of star clusters in different UVB models.
In this figure we use only the low-mass haloes that include one young star particle to focus on the scatter of $\fesc$, similarly to Figures~\ref{fig:distance} and \ref{fig:fesc_distance}.
We find that $\fesc$ steeply decreases with increasing local $n_{\rm H}$.  
When $n_{\rm H} \gtrsim 4$\,cm$^{-3}$, more than 90 per cent of the emitted ionizing photons cannot escape from the galaxy.
Once young stars born at such a high density region,
most of ionizing photons are absorbed on the spot because of high recombination rate.
Therefore we find that the variation of the local hydrogen number density also causes the large scatter in $\fesc$ for low-mass haloes.
This result is consistent with that shown in Figure~\ref{fig:fesc_distance}, because we expect lower $n_{\rm H}$ for grid cells at greater $\Rrel$.

\subsubsection{Dependence on Other Physical Quantities}

\begin{figure}
\includegraphics[scale=0.4]{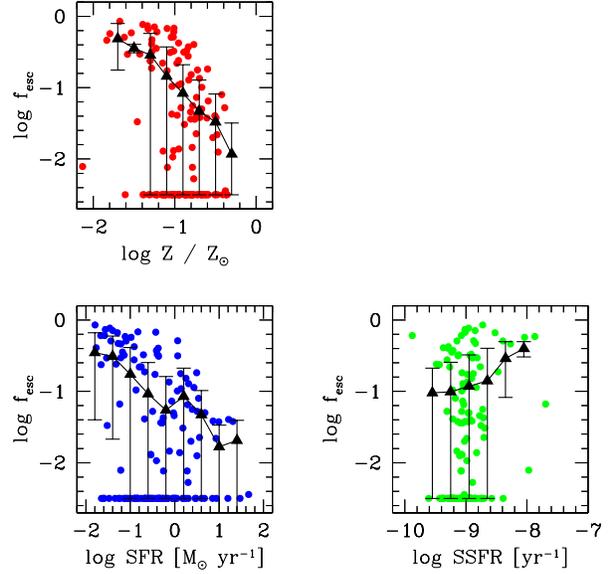}
\caption{
Escape fraction as a function of metallicity (upper panel), SFR (lower left panel), and specific SFR (lower right panel) for all the star-forming galaxies in the N144L10 Fiducial UVB run at $z=3$.
The negative correlation between $\fesc$ and metallicity is mainly caused by the positive correlation between halo mass and metallicity.  It is not the metals that directly controls $\fesc$. 
The data points with $\log \fesc < -2.5$ are set to $\log \fesc = -2.5$ for plotting purposes. 
}
\label{fig:other}
\end{figure}

We also examine the dependence of $\fesc$ on other physical quantities, 
such as metallicity, SFR, and specific SFR ($\equiv$\,SFR/stellar mass)
 in Figure~\ref{fig:other}.  This figure includes all star-forming haloes, just like Figure~\ref{fig:redshift}. 
We find that $\fesc$ decreases with increasing metallicity and SFR, 
and vice versa  for specific SFR.
These correlations are expected, because the metallicity and SFR are both positively correlated with galaxy stellar mass and halo mass. 
However, if the metallicity increases, the dust attenuation could have 
an extra effect on $\fesc$, which we will discuss in the next section.

%
%
\section{Discussion}
\label{sec:discussion}

\subsection{Dust Attenuation Effect}

\begin{figure}
\begin{center}
\includegraphics[scale=0.43]{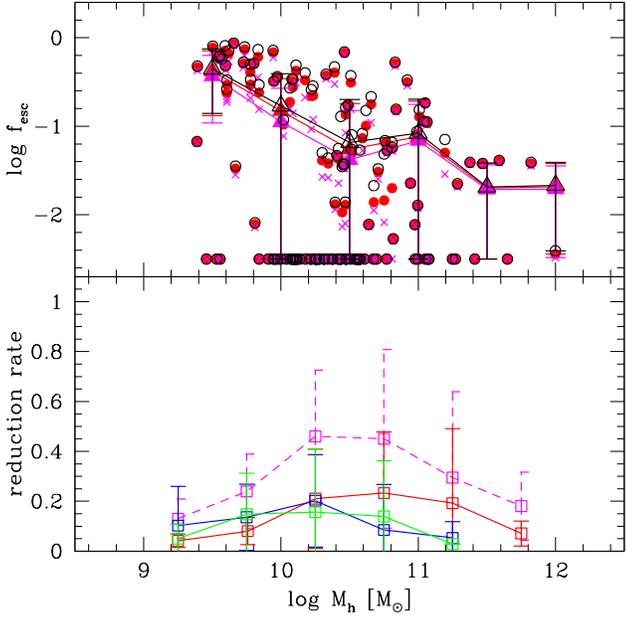}
\caption{
{\it Upper panel}: Comparison of $\fesc$ with and without dust extinction in the Fiducial N144L10 UVB model at $z=3$. Open black circles show $\fesc$ without dust extinction, and the filled red circles show $\fesc$ with dust extinction.  Magenta crosses show the case when the dust grain size range is changed to $0.03 - 0.3 \mu$m.  The triangles indicate the mean values in each mass bin with 1-$\sigma$ error bars.
The data points with $\log \fesc < -2.5$ are set to $\log \fesc = -2.5$ for plotting purposes. 
{\it Lower panel}: Mean reduction rate of $\fesc$ for each mass bin when including dust extinction.  Different colors show different redshifts (red: $z=3$, blue: $z=5$ and green: $z=6$).  
The dashed magenta result is when the dust grain size range is changed to $0.03 - 0.3 \mu$m.  The error bars are 1-$\sigma$. 
}
\label{fig:dust}
\end{center}
\end{figure}

Interstellar dust can decrease the $\fesc$ by absorbing the ionizing photons. 
We evaluate the effect of dust attenuation on $\fesc$ by comparing the RT 
result with and without the dust treatment, as described in \S~\ref{sec:dust}.
Figure~\ref{fig:dust} shows the values of $\fesc$ with and without the 
treatment of dust extinction, as a function of halo mass at $z=3$. 
The reduction rate of $\fesc$ does not depend on the halo masses very much, 
and ranges from 0 to 20 per cent, with an average of 14 per cent. 

If the dust-to-gas ratio is the same, the optical depth is roughly proportional to ${\bar a}_{\rm d}^{-1}$, where $\bar{a}_{\rm d}$ is typical dust size.
Note that $d \tau  = [ Q \pi a_{\rm d}^2 m_{\rm d} /  (4 \pi  a_{\rm d}^3  \rho /3)] d\ell \propto a_{\rm d}^{-1}$, where $m_{\rm d}$ and $\rho$ are dust mass and dust density. 
If we change the dust grain size distribution to a smaller size range of $0.03 - 0.3 ~\rm \mu m$, the mean reduction rate increases to $\sim 38$ per cent at $z=3$.  The weak dependence on halo mass is perhaps because most of the star-forming regions are enriched close to the solar metallicity, irrespective of the host halo masses. 

Our reduction rate is somewhat smaller than that reported by Y09, 
which is probably owing to the difference in the 
volume occupied by the metal rich gas.
In Y09, the model galaxy was an isolated system, and the ISM was 
globally mixed by the shock from supernova explosion,  because there were no further infall of pristine gas from intergalactic space. 
On the other hand, in the present work, pristine 
gas can accrete onto the haloes from intergalactic space, which reduces 
the volume fraction occupied by the metal rich gas around star-forming regions. 
However, the reduction rate of $\fesc$ reported in 
Y09 are still in the 1-$\sigma$ range of the present work. 

\citet{Razoumov10} and \citet{Gnedin08a} reported that the effect of dust attenuation on $\fesc$ is very small. 
In particular, \citet{Gnedin08a} argued that in their simulations, 
only the ionizing photons from stars in the outer disk can escape 
from the halo, and the escaping photons pass through only the 
low density regions with low metallicity and low dust content.  
This may lead to smaller variations of $\fesc$ in their simulations 
compared to other simulations that do not fully resolve the disk structure. 

\citet{Razoumov10} achieved a spatial resolution of $0.1 - 1$\,kpc 
in their SPH simulations using a zoom-resimulation technique, but it is not 
clear if they had a sufficient resolution to resolve the disk structure of 
galaxies.  Also, our SPH simulations have lower resolution than those of 
\citet{Razoumov10}, therefore we would expect a larger variation in $\fesc$ 
compared to theirs, and our results are consistent with this expectation. 
In our SPH simulations, the escaping photons may also pass through 
moderately high density regions with higher metal and dust content, 
with stronger effects of dust attenuation. 

Furthermore, our dust model is different from those of \citet{Gnedin08a} and \citet{Razoumov10}.  They adopted dust extinction curves of Large and Small Magellanic Clouds, whereas we use the dust size distribution of our Galaxy derived by \citet{Mathis77} for a silicate-type dust. 
Despite the fact that the effective absorption cross section of our dust model is smaller than in their models, 
the scatter of $\fesc$ in our simulations may be larger than theirs  
owing to the differences in both the resolution and dust treatment.    
In the future, we plan to improve our dust model by including 
the treatments of formation and destruction of dust particles.


\subsection{Contribution to IGM Ionization}

\begin{figure}
\includegraphics[scale=0.43]{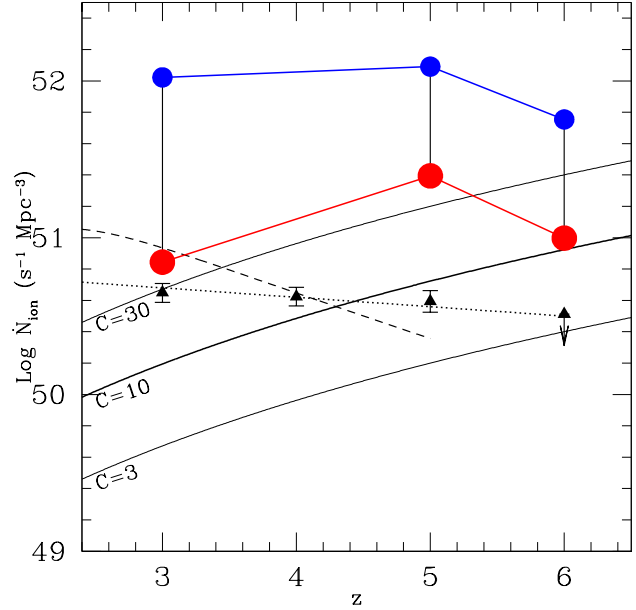}
\caption{
Comoving emission rate density of ionizing photons $\Nion$ as a function of redshift. Blue filled circles are for the  intrinsically radiated photons from all simulated galaxies, and the red filled circles are for the escaped ionizing photons after the RT calculation. 
Black solid lines show the required $\Nion$ to reionize the universe, derived by \citet{Madau99} for various clumping factors of IGM.
Black dotted line shows the contribution from QSOs estimated by \citet{Madau99}.
The filled squares are $\Nion$ derived from the Ly-$\alpha$ opacity data 
of IGM by \citet{Bolton07}.
}
\label{fig:reion}
\end{figure}

As a result of our RT calculation presented in the earlier sections, 
we are able to estimate the total number of ionizing photons that 
escape from all the star-forming galaxies in the entire simulation box. 
Our calculation includes haloes with $\Mh \gtrsim 10^{9.4} \Msun$ at $z=3$,  
and those with $\Mh \gtrsim 10^9 \Msun$ at $z=6$. 
Figure~\ref{fig:reion} compares the comoving emission rate density of 
ionizing photons $\Nion$ (i.e., the number of ionizing photons emitted
per unit time and per unit volume) in our N144L10 Fiducial UVB run to the 
required $\Nion$ to reionize the universe \citep{Madau99}, which is 
shown by the black solid curves for the clumping factors of 
$C=1, 3, 10,$ and 30.  
The blue solid circles are for the intrinsically radiated photons from 
all star-forming galaxies in our simulation, and the red circles are 
for the escaped photons after the RT calculation. 

The clumping factor of IGM at $z > 6$ is still very uncertain, and 
the results from numerical simulations vary depending on the resolution 
and the treatment of physical processes such as star formation and radiation 
transfer. Earlier, \citet{Gnedin97} suggested $C=30$ at $z \sim 6$, 
however \citet{Iliev07} reported $C=10$ using higher resolution simulations 
with a RT treatment. 
More recently, \citet{Pawlik09b} reported $C \sim 3 - 6$ using 
a cosmological SPH simulation with an optically thin approximation. 

Our results show that $\Nion$ of escaped ionizing photons after the RT 
calculation is greater than the required $\Nion$ to ionize the universe 
at $z=6$ if $C=10$, but below the required value if $C=30$. 
Therefore our fiducial simulation suggests that the star-forming galaxies 
can ionize the IGM as long as $C\le 10$. 

Our results on $\Nion$ is higher than those derived by \citet{Bolton07}, 
which was derived by using the results of cosmological SPH simulations and
observational data of Ly$\alpha$ forest.
Although the error bars of their data points look very small, 
it represents only the dispersion of Ly$\alpha$ opacity data.
There are still significant uncertainties in the spectral shape and
the mean free path of ionizing photons in \citet{Bolton07}.
In their calculation, they use the distance between Lyman limit systems 
as a mean free path of ionizing photons.
However many low density H\,{\sc i} gas clouds can decrease the mean
free path, and hence increase $\Nion$.
Together with the uncertainties in our simulation such as the resolution
and the details of star formation and feedback models, 
the differences between our results and that of \citet{Bolton07} 
can be accounted for. 

\begin{figure}
\includegraphics[scale=0.4]{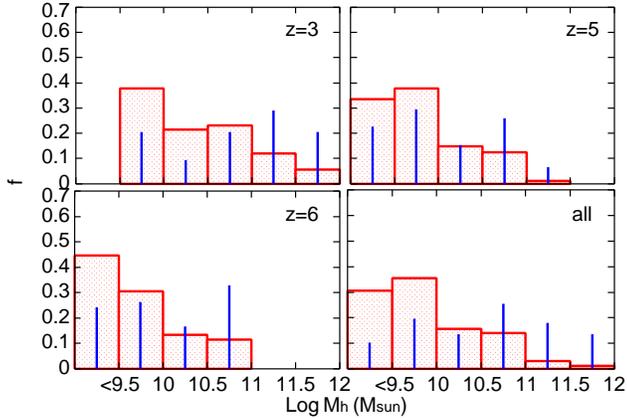}
\caption{
Number fraction of ionizing photons contributed by the haloes of different masses.  The blue lines are for the intrinsically radiated ionizing photons, and the red histograms are for the escaped ionizing photons after the RT calculation. 
This figure shows that most of the escaped ionizing photons mainly come from lower mass haloes. 
}
\label{fig:Nifrac}
\end{figure}

We also study the fractional contribution to the ionizing photons by 
the haloes with different masses, as shown in Figure~\ref{fig:Nifrac}. 
For the intrinsically radiated photons (blue lines), there is no clear 
trend with the halo mass, and all the haloes contribute roughly equal 
number of ionizing photons.  
However, the figure shows that most of the escaped ionizing photons 
(red histograms) come from the lower mass haloes ($\le 10^{10} \Msun$). 
This is because in our simulations, higher mass haloes have lower $\fesc$. 

Earlier in Section~\ref{sec:dependence}, we discussed the large variation 
of $\fesc$ derived by \citet{Shapley06}. 
Based on a clustering analysis, \citet{Adelberger05} reported that 
the sample in \citet{Shapley06} are hosted by haloes with $\Mh > 10^{11} \Msun$.  
When compared with our results in Figure~\ref{fig:Nifrac}, it suggests that
the sample of \citet{Shapley06} might not be tracing the bulk of ionizing 
sources at $z=3$, and only sampling the massive end of the distribution.

In addition, the detected sample in \citet{Iwata09} is brighter than 
$\sim 27$ mag in $R$ band.
If we use the equation in \citet{Madau98} and the relation between 
halo mass and SFR ($\Mh \sim SFR \times 10^{10} \Msun$),
the limiting magnitude of Iwata's sample corresponds to $\Mh \sim 1.5 \times 10^{10} \Msun$ (SFR $\sim 1.5 \Msun$\,yr$^{-1}$).
Therefore they are also tracing only the massive end of the distribution, 
although one order of magnitude deeper in halo mass than \citet{Shapley06}.  
Observations of fainter sources are 
needed to capture the entire ionizing radiation from lower mass haloes. 

In the SPH simulations used for this work, haloes with $\Mh \lesssim 10^{9} \Msun$ are not resolved well. For example, in the case of the Fiducial N144L10 run, the halo with $\Mh =2\times 10^9 h^{-1}\Msun$ consists of 100 dark matter particles. 
In the future, we plan to study $\fesc$ of even lower mass haloes using higher resolution simulations.  
If these low-mass haloes are the primary sources of ionizing photons at $z > 6$, the H\,{\sc ii} bubbles during the reionization epoch will be produced by numerous low mass haloes. 
In addition, the large dispersion in $\fesc$ that we found in this work suggests that the sizes of the H\,{\sc ii} bubbles may have a large variety in the 
early stages of reionization.


\subsection{Resolution Test}
\label{sec:resolution}

In the present work, we find that the value of $\fesc$ depends strongly on the distribution of star particles with respect to the high density gas.  
However, the clumpiness of ISM around the star-forming regions could strongly depend on  
the resolution limit of simulations, therefore 
it is important to evaluate the resolution effects on $\fesc$. 
In particular, the cosmological SPH simulations have a difficulty in resolving the clumpy structure of ISM when the number of SPH particles is very small, and the values of $\fesc$ for low-mass haloes may be more strongly affected by the limited resolution. 
When the clumpiness of ISM is very high, it may have both positive and negative effect on  
$\fesc$: the positive effect is that the ionizing photons may be able to escape through the void regions more easily, however those photons soon may be absorbed by the nearby high-density neutral clumps, which would be a negative effect. 

\begin{figure}
\includegraphics[scale=0.43]{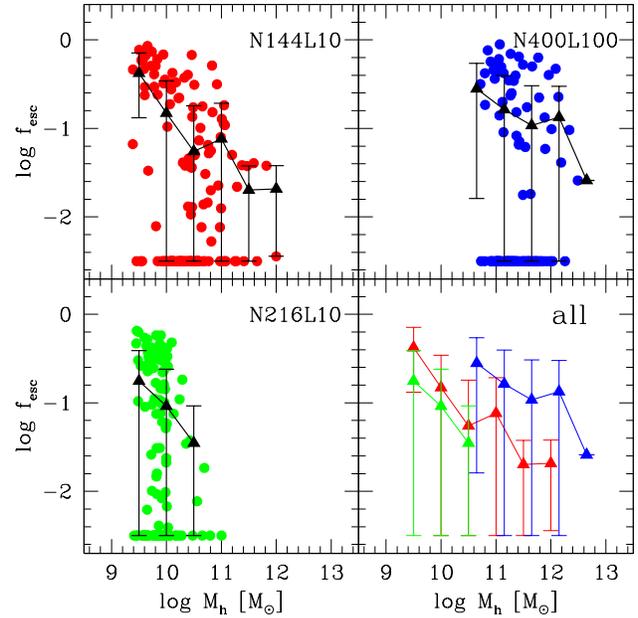}
\caption{
The resolution test on $\fesc$ using three runs with different resolution, as described in Table~\ref{table:sim}. 
The lower right panel compares the mean $\fesc$ in each mass bins for the three runs with 1-$\sigma$ error bars.
The points with $\log \fesc < -2.5$ are set to $\log \fesc = -2.5$ for plotting purposes.
}
\label{fig:resolution}
\end{figure}

To study the resolution effect, Figure~\ref{fig:resolution} compares $\fesc$ in the three runs with different resolution, as described in Table~\ref{table:sim}.
The lowest resolution run (N400L100) with a larger box size
shows overall higher $\fesc$ than the other runs.  
The highest resolution run (N216L10) gives similar values of $\fesc$ to the fiducial resolution run (N144L10), but somewhat lower $\fesc$ for the lower mass haloes with $\Mh \le 10^{9.5} \Msun$.  
Given the this result, it is quite possible that our $\fesc$ results have not converged yet, and we might still be overestimating $\fesc$ for the low-mass haloes in the present work. 
Nevertheless, it is true in all the runs that $\fesc$ has a large scatter for the low-mass haloes, and that $\fesc$ decreases on average with increasing halo masses. 

In Figure~\ref{fig:resolution}, for the N216L10 run, we had to confine the sample to the low-mass haloes with $\Mh < 10^{11}~\Msun$, because of the heavy computational load. In the N216L10 run, we need to set up a large grid of $> 500^{3}$ for the high-mass halos with $\Mh \sim 10^{12}~\Msun$, and these grids take too long to process with RT when we want to process a large sample. 
We will tackle the systematic study of haloes with higher resolution simulations using the next generation of supercomputers. 

For the low-mass haloes, the $\fesc$ of N216L10 run is smaller by a factor of $\sim 2$ than in the N144L10 run.  If we assume that $\fesc$ of all haloes becomes one half, the resulting $\Nion$ also becomes a half.
Then the red circles in Figure~\ref{fig:reion} would decrease by 0.3\,dex for the N216L10 run, and the threshold clumpiness factor for IGM reionization changes to $\sim 10(3)$ at $z=3(6)$.

\section{SUMMARY}
\label{sec:summary}

We have performed three-dimensional radiation transfer calculations of 
stellar radiation for a large number of high-$z$ star-forming galaxies 
in cosmological SPH simulations to explore the escape fraction of ionizing 
photons.  Our major findings are as follows:
\begin{itemize}
\item  The value of $\fesc$ decreases steeply with increasing halo mass, irrespective of numerical resolution. \\
\item There is a large dispersion in $\fesc$ for low-mass haloes with $\Mh \leq 10^{11} \Msun$. \\
\item  The values of $\fesc$ do not vary much with redshift and different UVB models. \\
\item  The average reduction rate of $\fesc$ owing to the dust attenuation effect is $\sim 14 \%$ with a large dispersion. \\
\item  The results of our Fiducial N144L10 run suggests that the star-forming galaxies can ionize the IGM at $z=3-6$, if the clumping factor is $C\lesssim 30$ (10) at $z=3$ (6). 
If we use the results of the N216L10 run, we roughly estimate that the above threshold values would change to $C\lesssim 10$ (3) at $z=3$ (6).
Our results suggest that the star-forming galaxies become the main contributor of IGM ionization at $3 \lesssim z \lesssim 6$.\\
\item The low mass haloes with $\Mh \lesssim 10^{10}\Msun$ are the main ionizing sources of IGM in our simulations owing to their high $\fesc$.  
The fraction of escaped ionizing photons coming from the haloes with $\Mh \le 10^{10} \Msun$ at $z=3-6$ is 70 per cent for the Fiducial N144L10 run.
\end{itemize}

As we summarised in Section~\ref{sec:intro}, the current results on the escape fraction of ionizing photons are confusing, as different results are obtained from different simulations. 
For example, \citet{Gnedin08a} argued that $\fesc$ increases with increasing halo mass in the range of $\Mh = 10^{10} - 10^{12}\Msun$, and their values of $\fesc$ were mostly less than a few per cent, much smaller than the other published work.  
The trend found in our simulations (decreasing $\fesc$ with increasing halo mass) is similar to that found by \citet{Razoumov10}, but we find lower $\fesc$ values despite of the fact that our simulations have lower resolution than their zoom-resimulations. 
Therefore the differences in $\fesc$ between our work and \citet{Razoumov10} cannot be explained simply by the resolution effect. 

We also note that \citet{Wise09} obtained much higher values of $\fesc$ ($\sim$0.4) than \citet{Gnedin08a} did, using the same AMR method, but for a different halo mass range.  Considering these facts, the differences that we see now in the results of $\fesc$ may have to do more with the different treatment of radiation transfer and the UV background radiation, rather than the resolution or numerical technique. 
However, more detailed comparisons are needed to make more definite statements. 

One of our main points is that the variation in $\fesc$ is caused by the different geometry of ISM distribution in the halo.
Recently \citet{Agertz10} suggested that supernovae feedback and 
star formation efficiency can determine the geometry of a disk galaxy 
\citep[see also][]{Sales10}.
Therefore the uncertainties in the treatment of star formation, feedback, and 
radiation transfer are all important for the calculations of $\fesc$, 
and we need to continue to improve these models through comparisons 
with future observations of high-$z$ galaxies. 

If we allow ourselves to speculate even further and combine all the current results mentioned above, it is possible that $\fesc$ has a peak at $\Mh \approx 10^9 - 10^{10} \Msun$ as a function of halo mass at $z=3-6$.  But this is highly speculative and by no means based on any definite physical arguments. 

The strength of our current work is the large sample size of galaxies that we processed with RT. Our simulations also adopt a new galactic wind model which produces more favorable results on the cosmic star formation rate and the IGM statistics such as C\,{\sc iv} mass density \citep{Choi10}. 
Although it is difficult to address the exact effect of our wind model on $\fesc$ unless we process simulations with different wind models with RT calculation, our test calculations showed that the effect is not so strong, and our main conclusions of this paper should remain unchanged even if we modify the wind model slightly. 
We consider that the most significant results in the current work are the large scatter of $\fesc$ for the low mass haloes, and its decline with the increasing halo mass.  The earlier works by other authors did not discuss the scatter among different haloes with a wide range of mass owing to their small sample size.

At $z \geq 7$, even lower mass haloes with $\Mh \lesssim 10^{9} \Msun$ may become the main sources of IGM ionization. However in such low-mass systems, 
the UV radiation of massive stars may influence the gas dynamics significantly \citep{Wise09}.  Simulations with higher resolution than presented in this paper are needed to follow the star formation in such low-mass systems, and we need to solve the hydrodynamics and radiation transfer simultaneously to examine the effect of radiative feedback.  
In the future, 
we plan to couple the RT with hydrodynamics and study the effects of radiative feedback by young stars and AGNs.

%
%
\section*{Acknowledgments}
HY thanks to M. Umemura for valuable discussions and comments. 
We are grateful to Volker Springel for providing us with 
the original version of GADGET-3, on which \citet{Choi09a} 
simulations are based.
This work is supported in part by the NSF grant AST-0807491, 
National Aeronautics and Space Administration under Grant/Cooperative 
Agreement No. NNX08AE57A issued by the Nevada NASA EPSCoR program, and 
the President's Infrastructure Award from UNLV. 
This research is also supported by the NSF through the TeraGrid resources 
provided by the Texas Advanced Computing Center (TACC).
Numerical simulations and analyses have been performed on the UNLV 
Cosmology Cluster, the {\it FIRST} simulator  and {\it T2K-Tsukuba} at 
Center for Computational Sciences, University of Tsukuba. 
KN is grateful to the hospitality of the IPMU, University of Tokyo, 
where part of this work was done.

%
%



\label{lastpage}

\end{document}